# Activation energy distribution of dynamical structural defects in RuO$_2$ films

Sheng-Shiuan Yeh,[1,2,*] Kuang Hong Gao,[1,3] Tsung-Lin Wu,[1] Ta-Kang Su,[1] and Juhn-Jong Lin[1,2,4,†]

[1]*Institute of Physics, National Chiao Tung University, Hsinchu 30010, Taiwan*
[2]*Center for Emergent Functional Matter Science, National Chiao Tung University, Hsinchu 30010, Taiwan*
[3]*Tianjin Key Laboratory of Low Dimensional Materials Physics and Preparing Technology, Department of Physics, Tianjin University, Tianjin 300072, China*
[4]*Department of Electrophysics, National Chiao Tung University, Hsinchu 30010, Taiwan*

Ruthenium dioxide (RuO$_2$) is an important metal widely used in nanoelectronic devices. It plays indispensable roles in the applications as catalyst and supercapacitors. A good understanding of the origin of the flicker or $1/f$ noise in RuO$_2$ will advance the design and efficiency of these applications. We demonstrate in a series of sputtered RuO$_2$ polycrystalline films that the $1/f$ noise originates from fluctuating oxygen vacancies which act as dynamical structural defects, i.e., moving scattering centers. Reducing the number of oxygen vacancies by adjusting thermal annealing conditions significantly reduces the noise magnitude $\gamma$, the Hooge parameter. We quantify the activation energy distribution function, $g(E)$, and calculate the oxygen vacancy density, $n_{\mathrm{TLS}}$, from the measured $\gamma$ value. We show that $g(E)$ can be explicitly expressed in terms of $\gamma(T)$ and the electronic parameters of the metal, where $T$ denotes temperature. The inferred $n_{\mathrm{TLS}}$ value is in line with the oxygen content determined from the x-ray photoelectron spectroscopy studies.

Corresponding authors, *Email: sunshine@nctu.edu.tw, †Email: jjlin@mail.nctu.edu.tw

**I. INTRODUCTION**

Transition metal oxides reveal a wide range of useful chemical, mechanical and electronic characteristics, and thus are of technological importance. Among them, ruthenium dioxide ($RuO_2$) attracts much attention due to the appealing properties of low electrical resistivity [1], high work function [2], as well as thermal and chemical stability under the ambient conditions [3]. It is a material widely used in the catalyst [4] and supercapacitor [5,6] applications. It has potential applications as interconnects in nanoelectronic devices [3]. Due to a high work function, it is used as gate electrodes to diminish leakage current through the dielectric layer. The dielectric thickness can therefore be reduced [7]. This is important for the downscaling of field-effect transistors [8] and dynamic random access memory devices [7]. The $RuO_2$-Au contacts have been utilized in radio-frequency microelectromechanical system (MEMS) switches, where an improved lifetime of 10 billion cycles was demonstrated [9]. Whereas it is a very useful material for nanoscale devices, a good understanding of the origin of the flicker noise in $RuO_2$ and how to possibly control the noise magnitude are still lacking. The undesirable $1/f$ noise, where $f$ denotes frequency, is deemed to hamper the performance of many electronic and spintronic devices [10].

The $1/f$ noise in conductors (metals and semiconductors) has been reviewed by, among others, Dutta and Horn [11], Weissman [12], and recently Fleetwood [13]. In conductors, the resistivity is $\rho = (n_e e \mu)^{-1}$, where $n_e$, $e$, and $\mu$ denote the charge carrier density, electronic charge, and carrier mobility, respectively. The relative resistivity fluctuation is given by the sum $-(\delta\rho/\rho) = \delta n_e/n_e + \delta\mu/\mu$, where the first term on the right-hand side of the equation denotes the charge carrier number fluctuations, and the second term denotes the charge carrier mobility fluctuations. In a Si metal-oxide-semiconductor field-effect transistor (MOSFET), the $1/f$ noise is well known to cause by $\delta n_e/n_e$, which microscopically originates from the trapping-detrapping processes of charge carriers in the near-interfacial SiO$_2$ dielectric [13]. On the other hand, because of the high carrier concentration [12], the $1/f$ noise in a metal is governed by $\delta\mu/\mu$, namely, the mobility fluctuations due to the charge carriers scattering off dynamical structural defects (or, moving scattering centers) [14-19]. In a particular material, which type/class of defect acts as the temporal "fluctuators" is usually difficult to specify [20]. Therefore, a physical, causal link between the $1/f$ noise and defect microstructure is often unknown.

In this work, we study the flicker noise and identify the responsible dynamic entities in metallic, polycrystalline RuO$_2$ films. We find that oxygen vacancies, which act as dynamical structural defects, play an important role in generating the $1/f$ noise [11,12]. A dynamical structural defect can be modeled as a two-level system (TLS) with an activation energy (or, potential barrier height) $E$, as schematically depicted in Fig. 1(a). The dynamical entity may be an atom or a group of atoms which switches back and forth between nearby metastable coordinate states [21,22]. Since a huge number of TLSs are present in a given disordered conductor, there must exist a wide distribution of the $E$ value, which can be described by an activation energy distribution function, $g(E)$. Despite decades of extensive investigations [11,12], this central quantity $g(E)$ which underlies the $1/f$ noise magnitude $\gamma$, the Hooge parameter, has only been determined to within an unspecified constant in literature. On the other hand, it is well established that $g(E)$ varies slowly over $k_B T$ [Fig. 1(d)], where $k_B$ is the Boltzmann constant and $T$ denotes temperature, and hence giving rise to the inverse frequency dependence of the noise power spectrum density (PSD) [23,24]. By combining the models of Dutta *et al*. [23] and Pelz and Clarke [19], we write down an explicit form of $g(E)$. This in turn allows us to compute the TLS density (in this case, oxygen vacancy density), $n_{\text{TLS}}$, which governs the measured $1/f$ noise magnitude in $RuO_2$. We show that, by adjusting thermal annealing conditions to modify the $n_{\text{TLS}}$ value, the $1/f$ noise amplitude can be changed by nearly two orders of magnitude. This observation is useful for the development of low-noise nanoelectronic devices and the improvement of catalysis efficiency of $RuO_2$ [4]. Compared to Si MOS devices [13], the underlying origins of the $1/f$ noise in metal films and nanostructures has received much less attention.

This paper is organized as follows. In Sec. II, we outline the theoretical concepts of the TLS mechanism of $1/f$ noise and write down an explicit expression for the activation energy distribution $g(E)$. In Sec. III, we describe our experimental method for sample fabrication, characterizations, and low-frequency noise measurement. Our results and discussion are given in Sec. IV. In particular, we provide evidence for a physical, causal link between the $1/f$ noise and oxygen vacancies in $RuO_2$, and quantitatively determine the function $g(E)$ and the quantity $n_{\text{TLS}}$. Section V contains our conclusion.

## II. THEORY

Microscopically, the scattering of conduction electrons off a single (say, the $i$th) TLS switching back and forth between coordinate states 1 and 2 [Fig. 1(a)] results in bi-state resistivity random telegraph noise (RTN), see a schematic in Fig. 1(b). For a thermal activation process (which is pertinent to our measurement temperature range of $90-350$ K), the relaxation rate is $\approx (1/\tau_D) e^{-E/k_B T}$, where the attempt frequency is taken to be the Debye frequency $1/\tau_D$. The resistivity noise PSD, $S_{\rho,i}$, for the RTN has a Lorentzian form [19,25]: $S_{\rho,i} = (\delta \rho_i)^2 (\tau_D/2) e^{E/k_B T} / \left[ 1 + \omega^2 (\tau_D/2)^2 e^{2E/k_B T} \right]$, where $\delta \rho_i$ is the resistivity difference between the two states, and $\omega = 2\pi f$ denotes angular frequency.

In a real sample, there naturally exists a large number of TLSs with a wide distribution of the relaxation time, owing to the broad distribution of $E$. For independently fluctuating TLSs, the total resistivity noise PSD of the sample is given by $S_\rho(\omega) = \int_0^\infty S_{\rho,i}(\omega, E) g(E) \Omega dE$, where the distribution function $g(E)$ is the number of TLSs per unit activation energy and per unit volume, and $\Omega$ denotes the sample volume. Figures 1(c) depicts that, at a given $\omega$ and a given $T$, $S_{\rho,i}(\omega, E)$ is a symmetric function of $E$ and peaks at $E = E_0(T) = -k_B T \ln(\omega \tau_D/2)$. The peak value is $S_{\rho,i}(\omega, E = E_0) = (\delta \rho_i)^2 / (2\omega)$ and the full width at half maximum (FWHM) is $\Delta E \simeq 2.6 k_B T$. Thus, the total noise PSD can be approximated by the definite integral $S_\rho(\omega) \simeq \int_{E_0 - \Delta E/2}^{E_0 + \Delta E/2} S_{\rho,i}(\omega, E) g(E) \Omega dE \simeq (\delta \rho_i)^2 \frac{\Omega}{2\omega} g(E_0) \times 2.6 k_B T$.

To facilitate comparison with experiment, it is conventionally written that [11,12] $S_\rho = \gamma \rho^2 / (N_c f^\alpha)$, where $\gamma$ is the dimensionless Hooge parameter, which characterizes the normalized magnitude of the flicker noise. $N_c = n_e \Omega$ is the total number of charge carriers in the sample, and the exponent $\alpha \simeq 1$. Therefore, the energy distribution can be rewritten as $g(E = E_0) \simeq \dfrac{4\pi}{2.6 k_B T} \dfrac{\gamma}{n_e \Omega^2} \dfrac{\rho^2}{(\delta \rho_i)^2}$.

Even after decades of extensive studies, $g(E)$ has only been treated as a *qualitative*, but not quantitative, function in literature [23]. The reason is that the resistivity variation $\delta \rho_i$ has previously been taken to be an unknown parameter. Here we would like to emphasize that $\delta \rho_i$ can be quantified. Based on the "local-interference" model of Pelz and Clarke [19], we have the good approximation $\delta \rho_i \simeq \dfrac{m v_F}{n_e e^2} \dfrac{\langle \sigma \rangle}{\Omega}$, because RuO$_2$ manifests vibrant TLS phenomena especially in small polycrystalline samples [26,27], where $m$ is the effective electron mass, $v_F$ is the Fermi velocity, and $\langle \sigma \rangle \simeq 4\pi / k_F^2$ is the averaged electron scattering cross section, with $k_F$ being the Fermi wavenumber. $g(E)$ can thus be explicitly expressed in terms of the known electronic parameters and the measured $\gamma(T)$ and $\rho(T)$ values of the metal, i.e.,

$$g\left(E = E_0(T)\right) \simeq \dfrac{4\pi \gamma(T) n_e}{2.6 k_B T} \left[ \dfrac{\rho(T) e^2}{m v_F \langle \sigma \rangle} \right]^2 . \qquad (1)$$

With $g\left(E = E_0(T)\right)$ being determined, the TLS number density $n_{\text{TLS}}$ can be directly calculated from the equation

$$n_{\text{TLS}}(T) \simeq g(E = E_0) \times 2.6 k_B T = 4\pi \gamma(T) n_e \left[ \dfrac{\rho(T) e^2}{m v_F \langle \sigma \rangle} \right]^2 . \qquad (2)$$

That is, $n_{\text{TLS}}(T)$ is given by the product of $g\left(E = E_0(T)\right)$ and the FWHM of the Lorentzian function, and the number density $n_{\text{TLS}}(T) \propto \gamma(T)$.

## III. EXPERIMENTAL METHOD

RuO$_2$ thin films were deposited by dc magnetron sputtering on 300-nm-thick SiO$_2$ capped Si substrates. Before deposition, the substrates were pre-patterned with a multiple electrode configuration for the low-frequency noise measurements, using the standard electron-beam lithography technique. A typical sample geometry is shown in Fig. 2. A commercial RuO$_2$ target (99.95% purity, Admat Midas Inc., Dover, Delaware, USA) was used for the sputtering deposition. The base pressure of the vacuum chamber was $\approx 3.0 \times 10^{-6}$ torr, and the sputtering was carried out in an atmosphere of argon and oxygen mixture of $\approx 3.0 \times 10^{-2}$ torr. The sample stage was held at room temperature. During sputtering, the Ar and O$_2$ flow rates were kept at 4.2 and 2.0 sccm, respectively. The sputtering power was 100 W, and the deposition rate was $\approx 10$ nm/min. The thickness of the deposited films was 60, 100 or 150 nm. After deposition, the films were first annealed for 10 min in an O$_2$ atmosphere at $500\,°C$. This annealing process effectively removed any residual Ru phase in the as-deposited film. To study the effect of oxygen vacancies on the $1/f$ noise magnitude, a series of films were further annealed for another 5 or 15 minutes in O$_2$, also at $500\,°C$. This additional annealing process significantly reduced the number of oxygen vacancies in the sample. After the final O$_2$ annealing, some films were subsequently annealed in an Ar gas at $500\,°C$ to reintroduce oxygen vacancies into the sample [28,29]. For the x-ray diffraction (XRD) studies of the crystal structure and the x-ray photoelectron spectroscopy (XPS) studies of the oxygen contents, un-patterned films on glass substrates were deposited simultaneously for each batch of samples.

The scanning electron microscope (SEM) image in Fig. 2 shows a representative sample with eight electrodes labeled from a to h. The sample was 1 $\mu$m wide and the distance between any two neighboring electrodes was 10 $\mu$m. The Cr/Au (10/100 nm) electrodes on each sample were deposited via thermal evaporation after the final $O_2$ or Ar annealing process was done. It should be stressed that the contact resistance between the sample and each electrode was very small, typically $\approx 0.2$ $\Omega$. The low-frequency noise were measured using the ac bridge technique [30,31]. The measurement circuit is schematically depicted in Fig. 2. A lock-in amplifier (Stanford Research Systems Model SR830), a preamplifier (Stanford Research systems Model SR560), and a dynamic signal analyzer (Stanford Research Systems Model SR785) were used. The temporal voltage fluctuations were measured by SR785, and the voltage noise PSD $S_V$ was calculated by LabVIEW programs. The sampling rate of SR785 was set to be 1024 Hz. To minimize the measured noise uncertainties, for a given sample bias voltage $V$, $S_V$ was measured 10 times, each time taking 30 s or about 30,000 readings. The $S_V$ signals were then averaged. The processes were repeated for seven bias $V$ values for each measurement temperature.

The resistivities $\rho(300\ \mathrm{K})$ for the 150-nm-thick films annealed in $O_2$ for 10, 15 and 25 min were 492, 479, and 371 $\mu\Omega$ cm, respectively. For 25 min annealing in $O_2$ followed by 5 min annealing in Ar, $\rho(300\ \mathrm{K}) = 490\ \mu\Omega$ cm. For the 100-nm-thick films annealed in $O_2$ for 10 and 15 min, $\rho(300\ \mathrm{K}) = 470$ and 247 $\mu\Omega$ cm, respectively. For 10 min annealing in $O_2$ followed by 5 min annealing in Ar, $\rho(300\ \mathrm{K}) = 389\ \mu\Omega$ cm. Since these films were fairly disordered, their resistivities depended only weakly on temperature. In each film, the resistivity decreased by $\leq 5\%$ as $T$ was reduced from 300 K to 90 K.

## IV. RESULTS AND DISCUSSION

### A. X-ray diffraction and x-ray photoelectron spectroscopy studies

The XRD patterns for three representative films with different thicknesses and annealed for 10 min in $O_2$ are shown in Fig. 3(a). This figure indicates that all films are polycrystalline with diffraction patterns characteristic of the $RuO_2$ rutile structure. No signature of a residual Ru phase is seen. From the positions of the (110) and (101) peaks, the lattice constants $a$ and $c$ are calculated to be 4.53 and 3.06 Å, respectively, in good agreement with recorded values [32]. From the FWHM of the (110) peak, the average grain size is calculated through the Scherrer equation [33,34] to be $\approx 6$ nm *in all samples* studied in this work, regardless of the differing thermal annealing conditions. Figure 3(b) shows the (110) peak in an enlarged scale for four 150-nm-thick films annealed under various conditions, as indicated. This figure clearly illuminates that the FWHM remains the same for all films.

We have carried out XPS studies to determine the O contents in our films. The main panel and the inset of Fig. 4 show respectively the O 1s and the Ru $3p_{3/2}$ XPS spectra for a 150-nm-thick film annealed for 10 min in $O_2$. The asymmetric peak shape of the O 1s spectra suggests the presence of more than one oxide species in the sample. Quantitatively, this spectra can be deconvoluted into three peaks (blue Gaussian-Lorentzian curves) with binding energies of 529.8, 531.3, and 532.8 eV. The red curve is the sum of the three blue curves, which well reproduces the experimental result. These three peak values are in good consistency with those values reported in literatures [35], and are known to originate from the lattice $O^{2-}$, dissolved O atoms, and chemisorbed hydroperoxy radicals on the surface, respectively. The Ru $3p_{3/2}$ spectra illustrate a peak value at 463.0 eV, in good agreement with previous result [36]. The oxidation state $-2$ of the lattice O suggests that our films possess the $RuO_2$ phase, confirming the XRD results (Fig. 3).

The relative atomic ratio of O to Ru in the $RuO_x$ rutile lattice can be estimated through the relation $x = n(O)/n(Ru) = A(O)S(Ru)/A(Ru)S(O)$ (Ref. [37]), where $n(O)$ [$n(Ru)$] denotes the atomic density of O (Ru), $A(O)$ [$A(Ru)$] denotes the XPS peak area of O 1s lattice (Ru $3p_{3/2}$), and $S(O) = 2.93$ [$S(Ru) = 6.78$] denotes the elemental sensitivity factor of O 1s (Ru $3p_{3/2}$). From the data in Fig. 4, we obtain the oxygen content $x \approx 1.923$. Similarly, we obtain $x \approx 1.938$, $1.983$ and $1.960$ for the films annealed for 15 min in $O_2$, 25 min in $O_2$, and 25 min in $O_2$ followed by 5 min in Ar, respectively. These four $x$ values give rise to the ratios of the number of O vacancies to the number of total O atoms (in a $RuO_2$ unit cell) in our films to be $n_{O,V}/n_{O,total} \approx 3.8\%$, $3.1\%$, $0.85\%$ and $2.0\%$, respectively. These results clearly indicate that lengthening thermal annealing in an $O_2$ gas progressively reduces the O vacancy density, while thermal annealing in Ar effectively removes O atoms and generates O vacancies.

### B. Excess low-frequency noise

In practice, to study the $1/f$ noise in a metal, one applies a small current $I$ to measure the voltage noise PSD which is given by the Hooge formula [24]

$$S_V = \frac{\gamma V^2}{N_C f^\alpha} + S_V^0, \quad (3)$$

where $V = 2V_{rms}$ for the ac bridge measurement configuration depicted in Fig. 2, with $V_{rms}$ being the root-mean-square voltage drop across one-half length of the sample [30]. $S_V^0$ is the PSD of the background noise, which contains the Johnson-Nyquist noise of the sample and the input noise of the preamplifier used in the circuit. In this study, $S_V^0$ is negligibly small compared with the sample noise, see below. For an ohmic conductor, $S_V/V^2 = S_\rho/\rho^2$. Figure 5(a) shows the PSD $S_V$ for a 100-nm-thick RuO$_2$ film annealed for 10 min in O$_2$, where the voltage noise was measured between the electrodes b and d, see Fig. 2. Each curve in Fig. 5(a) represents the averaged $S_V$ measured over 10 times, as described in Sec. III. The background noise PSD (bias voltage $V = 0$) is $S_V^0 \approx 1 \times 10^{-16}$ V$^2$/Hz. Upon the application of a finite bias $V$, we see the variation $S_V \propto 1/f$ below ~ 2 Hz, and the $S_V(f)$ amplitude increases with increasing $V$.

Empirically, to obtain an accurate $\gamma$ value for a given sample, we write $\langle fS_V \rangle = \langle \gamma V^2/N_C \rangle + \langle fS_V^0 \rangle = \gamma V^2/N_C + \langle f \rangle S_V^0$ from Eq. (3), where $\langle fS_V \rangle$ denotes the average of the product of each discrete $f_i$ and $S_{Vi}$ readings in the data set, and $\langle f \rangle$ denotes the average of $f_i$ in the $S_V \propto 1/f$ inverse frequency regime. The slope $\gamma/N_c$ of the $\langle fS_V \rangle$ versus $V^2$ plot immediately gives the $\gamma$ value, where $N_C$ is known. For RuO$_2$, we estimate $n_e \simeq 5 \times 10^{28}$ m$^{-3}$, $k_F \simeq 1 \times 10^{10}$ m$^{-1}$, and $v_F \simeq 8.2 \times 10^5$ m/s through the free-electron model [27]. Note that extracting $\gamma$ value by this method excludes any possible contribution from the background noise $S_V^0$. Figure 5(b) shows $\langle f S_V \rangle$ as a function of $V^2$, where a linear dependence is evident (black squares). The straight black solid line is a least-squares fit. This behavior suggests that the measured $S_V$ must originate from the resistance fluctuations. Moreover, we see that the slope in Fig. 5(b) is reduced by a factor of $\simeq 2$ as the sample length is doubled (blue triangles, which were measured between the electrodes e and g). Thus, the contribution of the contact noise from the Cr/Au electrode/sample interfaces must be negligible, compare with the intrinsic sample noise.

To unravel the origin of the $1/f$ noise in RuO$_2$, we have measured the low-frequency noise in a series of samples underwent various thermal annealing conditions. The annealing conditions were adjusted to modify the number of oxygen vacancies in the samples, as confirmed by the XPS studies (Fig. 4). In particular, we have measured the temperature dependence of $\gamma$ in order to extract the distribution function $g(E = E_0)$, Eq. (1). Figure 6(a) shows $\gamma$ as a function of $T$ for four 150-nm-thick films subject to differing annealing conditions, as indicated. This figure reveals that $\gamma$ has a (local) maximum value around 300 K. (Note that the ordinates in Fig. 6 are plotted in logarithmic scales.) Below 300 K, $\gamma$ decreases with decreasing $T$. Compared with that in the film annealed for 10 min in O$_2$ (black symbols), the $\gamma(300 \text{ K})$ value in the film annealed for 25 min in O$_2$ (blue symbols) decreases by nearly two orders of magnitude, from $1.3 \times 10^{-2}$ to $3.1 \times 10^{-4}$. With an additional annealing for 5 min in Ar for the 25 min annealed film, the $\gamma(300 \text{ K})$ value (pink symbols) is increased by about one order of magnitude, from $3.1 \times 10^{-4}$ to $2.1\times 10^{-3}$. It should be emphasized that the variation in the $\gamma$ value is not due to a change in the grain size [30], because the grain size remains essentially the same ($\approx 6$ nm) for all thermal annealing conditions, as indicated in Fig. 3(b). (Atom diffusion along the grain boundary may cause flicker noise [38].) The large changes in the $\gamma$ value seen in Fig. 6(a) strongly suggest that oxygen vacancies in RuO$_2$ films must play an important role in generating the $1/f$ noise. This can be explained in terms of the schematic depicted in Fig. 1(a), where a TLS is formed by an oxygen vacancy and nearby oxygen atom(s). An oxygen atom can switch between the two coordinate states separated by potential barrier height *E*. Annealing in an O$_2$ gas largely reduces the density of oxygen vacancies (equivalently, reducing $n_{\text{TLS}}$), and hence significantly decreases the $\gamma$ value. On the contrary, a thermal annealing process in an Ar gas generates a large amount of oxygen vacancies (equivalently, increasing $n_{\text{TLS}}$), leading to a significant increase in the $\gamma$ value. For comparison, we note that $\gamma \approx 10^{-4}-10^{-2}$ in typical metals [39].

To further confirm that oxygen vacancies are the responsible dynamic entities, we have also studied three 100-nm-thick films. Two of the films were annealed in O$_2$ for 10 and 15 min, respectively. The third film was first annealed in O$_2$ for 10 min, followed by another annealing in Ar for 5 min. As shown in Fig. 6(b), a marked increase in the $\gamma$ value after Ar annealing is evident (blue symbols). Therefore, we conclude that oxygen vacancies and their formation of TLSs cause the $1/f$ noise in RuO$_2$ polycrystalline films. (For the films annealed for 10 and 15 min in O$_2$, the $\gamma$ value in the 100-nm-thick films is smaller by a factor of ~4 than that in the corresponding 150-nm-thick films. This can be ascribed to the fact that thermal annealing is more effective for thinner films.)

In the $1/f$ noise studies, it is highly desirable to learn the role of the dynamic defect in a particular material system. As is well-known, in Si MOSFETs, the oxygen vacancies in the near-interfacial $SiO_2$ form traps in the dielectrics. These traps cause charge carrier number fluctuations, leading to flicker noise [40-42]. In contrast, in $RuO_2$ films and the high-temperature superconductor $YBa_2Cu_3O_{7-\delta}$ [43,44], the TLSs formed by oxygen vacancies result in mobility fluctuations, producing the $1/f$ noise. It is interesting to remark that in $RuO_2$ which has a Fermi energy of a few eV [45], if any oxygen vacancies should form vacancy-related traps with level energies lying below the conduction band minimum, the detrapping time would be extremely long [$\geq (10^{15}-10^{17})$ s at 300 K, assuming an attempt-to-escape frequency of $\sim (10^{11}-10^{13})$ s$^{-1}$ (Ref. [46])]. This long detrapping time definitely cannot cause carrier number fluctuations to generate the low-frequency noise shown in Fig. 5(a).

### C. Activation energy distribution

From the measured $\gamma(T)$ values and Eq. (1), we have calculated the distribution function $g(E)$. Figures 7(a) and 7(b) show the variation of $g(E)$ with activation energy $E = -k_B T \ln(\omega \tau_D / 2)$ for several 150- and 100-nm-thick films underwent various annealing conditions, as indicated. We note that the $g(E)$ function is quantitatively determined, subject to no undefined parameters. The activation energy $E$ characteristic of those predominant TLSs which are probed between our measurement temperature 90−350 K varies from 0.25 to 0.95 eV. These $E$ values are here calculated by using a Debye frequency $1/\tau_D = 5\times 10^{13}$ Hz (or, Debye temperature $\theta_D = 400$ K) [47] and for the representative frequency of $f = 1$ Hz. Provided that the local-interference model is valid, one may measure $\gamma$ at even higher $T$ to obtain $g(E)$ distribution at higher $E$ values, because $E_0 \propto T$. At low temperatures, however, one will need to treat the universal conductance fluctuation (UCF) model for the $1/f$ noise [26]. Inspection of Fig. 7 indicates that $g(E)$ is relatively flat over the scale of $k_B T$, as expected [11,12]. At 300 K, $g(E)$ takes a value on the order from $10^{27}$ to $10^{29}$ eV$^{-1}$ m$^{-3}$, depending on the thermal annealing conditions.

Intuitively, for a perfect RuO$_2$ single crystal without any lattice randomness or inhomogeneities, the activation energy should take a fixed value determined by the lattice site(s) of the responsible oxygen atom/vacancy. However, in a polycrystalline film, lattice distortion and disorder may occur near the grain boundaries, leading to a distribution of $E$. Because the grain size ($\approx 6$ nm) is small in our films, a large portion of the atoms in a grain must reside near grain boundaries. Depending on their distance to a grain boundary and the characteristics (e.g., strains, stresses, or softness) of the grain boundary, different degrees of lattice distortion may take place. Thus, the $E$ value can have a distribution. This explains our observation of the distribution function $g(E)$ in Fig. 7. Previously, the inhomogeneity-induced distributions of $E$ were observed in Ag films and Cu whiskers and films [11,23].

By plotting the ordinates in linear scales, we clearly see in Fig. 7(c) that $g(E)$ demonstrates a peak at $E \approx 0.8$ eV. This feature is meaningful. According to recent theoretical density functional calculations, the activation energy of an on-top oxygen atom to the nearby bridging-oxygen vacancy is estimated to be ~ 0.7 eV in $RuO_2$ [48,49]. This theoretical estimate is reasonably close to our peak value of ~ 0.8 eV, thus can be supportive of the predominant role of oxygen vacancies as dynamical structural defects. We should mention that the present study does not exclude the possible existence of other local maxima in the $g(E)$ distribution. To clarify this issue requires measurements over a large temperature range.

A good understanding and a fair control of the $g(E)$ distribution are desirable for fostering the performance of nanoscale devices. Good knowledge about $g(E)$ can help to improve the catalysis efficiency of $RuO_2$ [48], as mentioned. TLSs can also cause dissipations and hamper the quality factors in nanoelectromechanical systems (NEMS) [50]. Besides, TLSs can generate frequency noise, namely, random fluctuations of the resonance frequency, in a small mechanical resonator. The frequency noise PSD was found to depend linearly on $g(E)$ [51,52]. The resonance frequency fluctuations will limit the resolution for tiny mass (force) detection [53]. Quantitative information about $g(E)$ is clearly crucial for these applications. To our knowledge, no other experimental method can offer a *quantitative* evaluation of $g(E)$ like that presented in Fig. 7.

### D. Estimate of the number of oxygen vacancies

With the known value of $g(E)$, the TLS (oxygen vacancy) density $n_{TLS}(T)$ can be calculated from Eq. (2). We found that, for the 150-nm-thick films annealed in $O_2$, $n_{TLS}$ ($\approx 1 \times 10^{26}$ $m^{-3}$ at 300 K) is reduced by nearly two orders of magnitude in the 25 min annealed film, compared with that ($\approx 7\times10^{27}$ m$^{-3}$ at 300 K) in the 10 min annealed film. This corresponds to a decrease of the ratio of the number of oxygen vacancies to the number of total oxygen atoms, $n_{O,v}/n_{O,total}$, from $\approx 10\%$ down to $\approx 0.2\%$ in the RuO$_2$ rutile structure. The additional annealing in Ar for 5 min increases $n_{TLS}$ ($\approx 1\times10^{27}$ m$^{-3}$ at 300 K) by one order of magnitude, corresponding to an increase in $n_{O,v}/n_{O,total}$ from $\approx 0.2\%$ to $\approx 2\%$. As the temperature is reduced to 100 K, we obtain $n_{TLS} \approx 1\times10^{25}$ m$^{-3}$ in the 150-nm-thick film annealed for 25 min in O$_2$, corresponding to the ratio of $n_{O,v}/n_{O,total} \approx 0.02\%$. Similar results are obtained for the 100-nm-thick films.

It is instructive to compare these $n_{TLS}$ values inferred from the $1/f$ noise measurement with those corresponding values determined from the XPS technique. As an example, for the 150-nm-thick film annealed for 10 min in O$_2$, the $n_{TLS}$ value inferred from Eq. (2) is a factor of 2.6 larger than the O vacancy density ($n_{O,v}/n_{O,total} \approx 3.8\%$) extracted from the XPS spectra discussed in Sec. IV.A. We consider this level of discrepancy as satisfactory, which can (partly) be ascribed to the uncertainties in our evaluations of the electronic parameters for RuO$_2$. For instance, a 20% uncertainty (decrease) in our estimate of $n_e$ will result in a factor of $\approx 2$ decrease in the calculated $n_{TLS}$ value, according to Eqs. (2) and (3). Another possible origin for the discrepancy is that, with the generation of an oxygen vacancy, the nearby oxygen atoms may be attracted toward the vacancy, distorting the bonding with the neighboring oxygen atoms. Such distortions may possibly induce additional TLSs, giving rise to a higher $n_{TLS}$ ($> n_{O,v}$) value. This issue deserves further study.

## V. CONCLUSION

We have studied the excess low-frequency noise in metallic, polycrystalline RuO$_2$ films. We find that the $1/f$ noise is originated from oxygen vacancies which act as dynamical structural defects. The $1/f$ noise magnitude can be reduced or increased by thermal annealing in an O$_2$ or Ar atmosphere. We quantitatively extract the activation energy distribution function $g(E)$ and the TLS (oxygen vacancy) number density $n_{\text{TLS}}$. Our results can be useful for advancing the performance of the various applications involving RuO$_2$. Our finding may help to shed light on the properties of the TLSs which naturally form in the heavily used aluminum-oxide dielectric. These TLSs in AlO$_x$ will likely be the ultimate source(s) for the quantum decoherence in superconducting qubits [54], and have recently been intensely investigated [20].

## ACKNOWLEDGEMENTS

We thank S. Kirchner and J. Kroha for valuable discussion on the dynamical structural defects and activation energy distribution in the RuO$_2$ rutile structure. We also thank I. K. Cheng for the XPS measurements and discussion. This work was supported by Ministry of Science and Technology through Grant numbers MOST-106-2112-M-009-007-MY4 and MOST-107-3017-F009-003, and the Center for Emergent Functional Matter Science of National Chiao Tung University from The Featured Areas Research Center Program within the framework of the Higher Education Sprout Project by the Ministry of Education (MOE) in Taiwan.

**Figures and Captions**

**Figure 1**

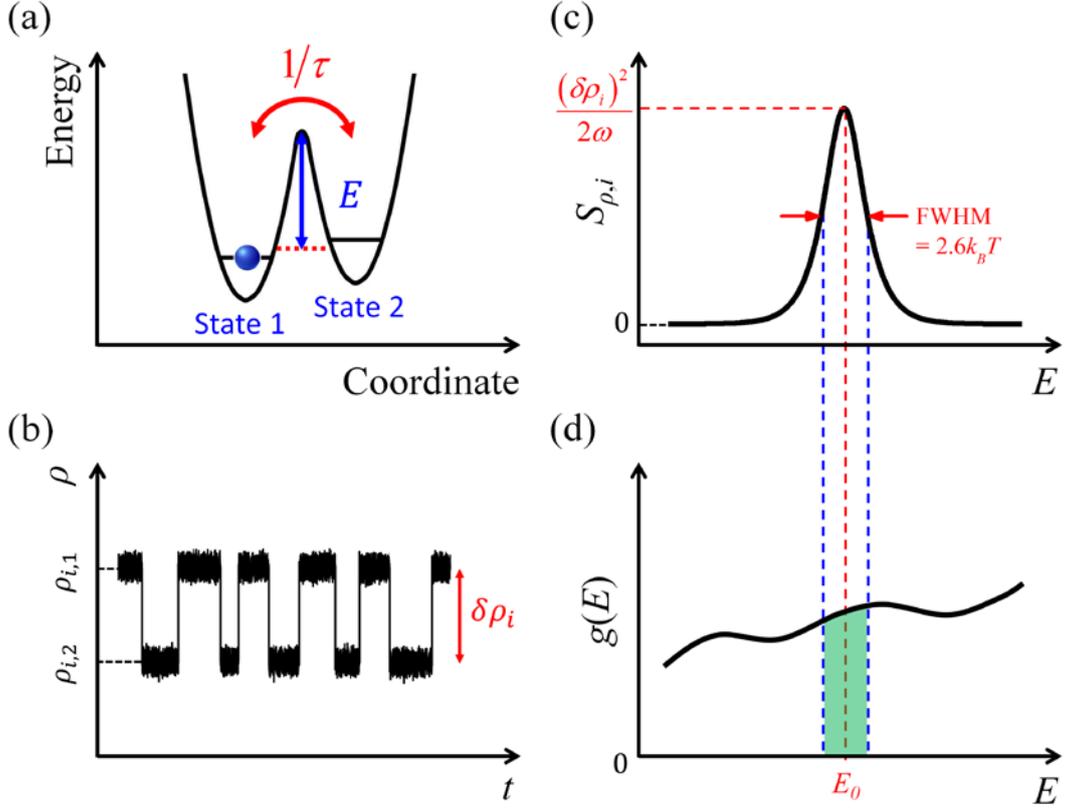

**FIG. 1.** (a) A schematic diagram for a (the *i*th) two-level system (TLS) with a double-well potential and activation energy $E$. $1/\tau \approx (1/\tau_D)e^{-E/k_BT}$ is the thermally activated switching rate between coordinate state 1 and state 2. (b) Random telegraph noise caused by the *i*th TLS. The variation between the upper ($\rho_{i,1}$) and lower ($\rho_{i,2}$) resistivity states is $\delta\rho_i = \rho_{i,1} - \rho_{i,2}$. (c) Resistivity noise power spectrum density (PSD) $S_{\rho_i}(E)$ due to a (the *i*th) TLS at measurement temperature $T$ for a given angular frequency $\omega$. $S_{\rho_i}(E)$ centers at $E_0 = -k_BT\ln(\omega\tau_D/2)$, with a peak value of $(\delta\rho_i)^2/(2\omega)$ and the full width at half maximum (FWHM) of $2.6k_BT$. (d) A schematic diagram depicting the distribution function $g(E)$ varying slowly relative to $k_BT$.

**Figure 2**

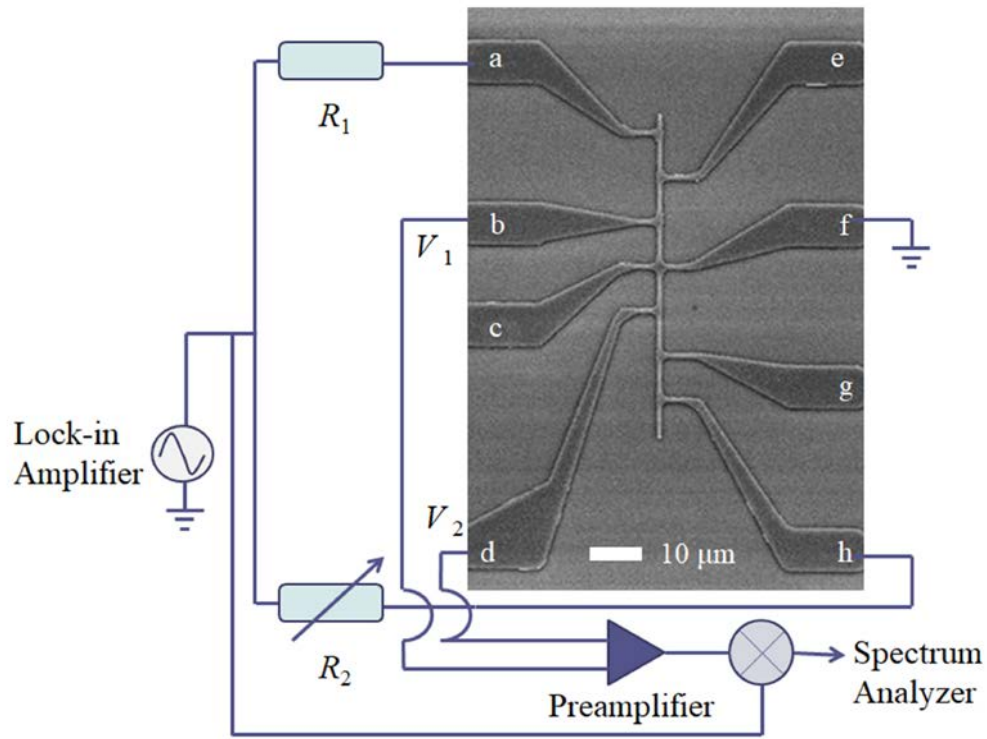

**FIG. 2.** SEM image of a patterned $RuO_2$ film, with a schematic low-frequency noise measurement circuit. The eight Cr/Au electrodes are labeled from a to h. The ballast resistor $R_1$ is usually set at a value that is at least 10 times larger than the sample resistance. The adjustable resistor $R_2$ is used to balance the bridge. The $V_1$ and $V_2$ electrodes can be reconnected to measure different segments of sample.

**Figure 3**

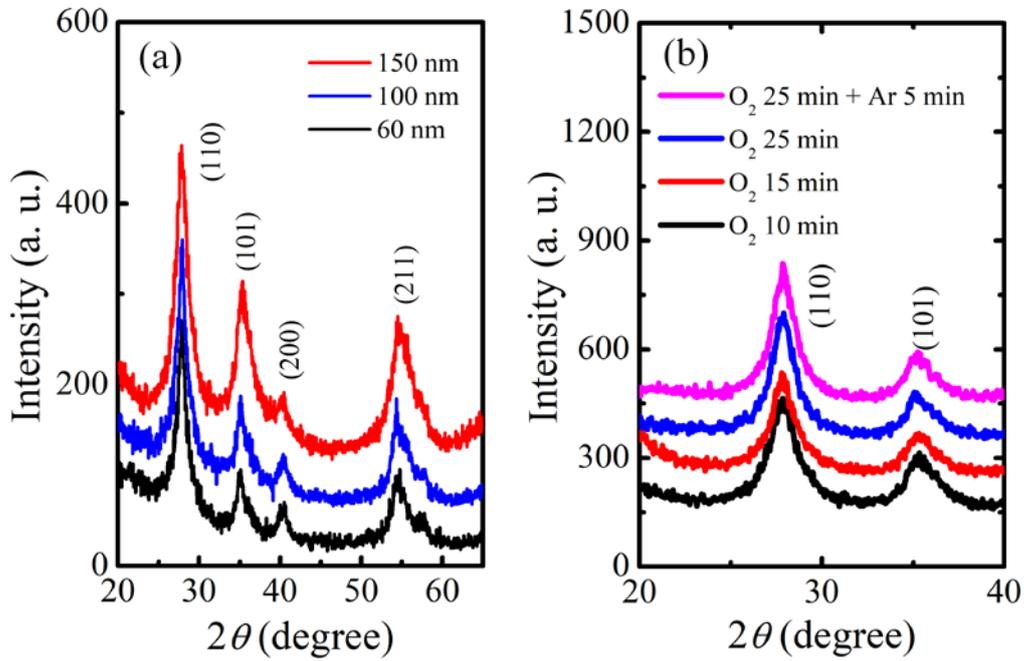

**FIG. 3.** (a) XRD patterns of three RuO$_2$ films annealed for 10 min in O$_2$ and with thickness of 60, 100, and 150 nm, as indicated. (b) XRD patterns for the (110) peak for four 150-nm-thick RuO$_2$ films annealed under various conditions, as indicated. Note that the peak width remains essentially the same, regardless of the thermal annealing processes.

**Figure 4**

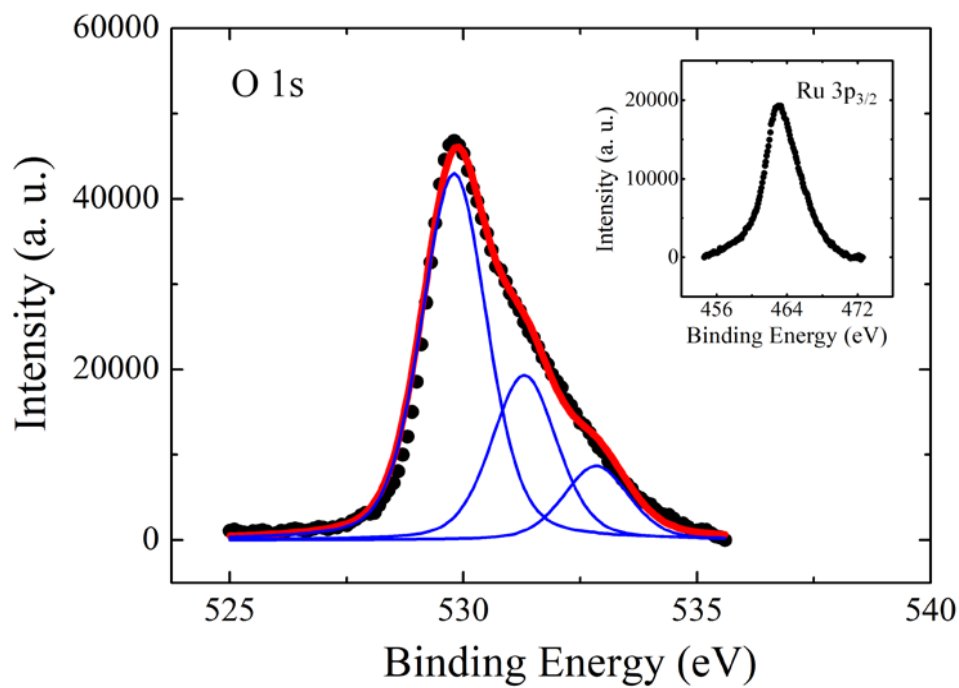

**FIG. 4.** The O 1s XPS spectra of a 150-nm-thick $RuO_2$ film annealed for 10 min in $O_2$. The spectra can be quantitatively deconvoluted into three peaks (blue Gaussian-Lorentzian curves) with binding energies of 529.8, 531.3, and 532.8 eV. The red curve is the sum of the three blue curves. Inset: the Ru $3p_{3/2}$ XPS spectra of the same film. The peak value indicates a binding energy of 463.0 eV.

**Figure 5**

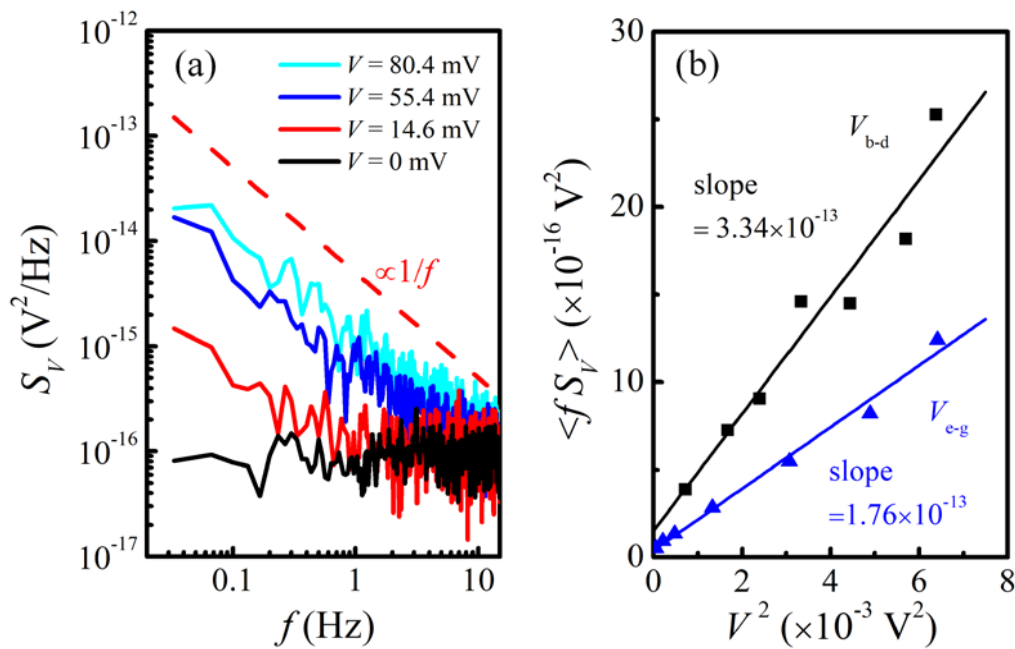

**FIG. 5.** (a) Voltage noise power spectrum density (PSD) as a function of frequency for a 100-nm-thick $RuO_2$ film annealed for 10 min in $O_2$ and measured at 296 K. The PSD magnitude increases with increasing bias voltage $V$, as indicated. The dashed line is drawn proportional to $1/f$, and is a guide to the eye. (b) Variation of $\langle f S_V \rangle$ with $V^2$ for the sample shown in (a). Black squares (blue triangles) are the data measured between the electrodes b and d (e and g), see the electrode configuration in Fig. 2. The solid straight lines are linear fits. The two slopes differ by a factor of 1.9.

**Figure 6**

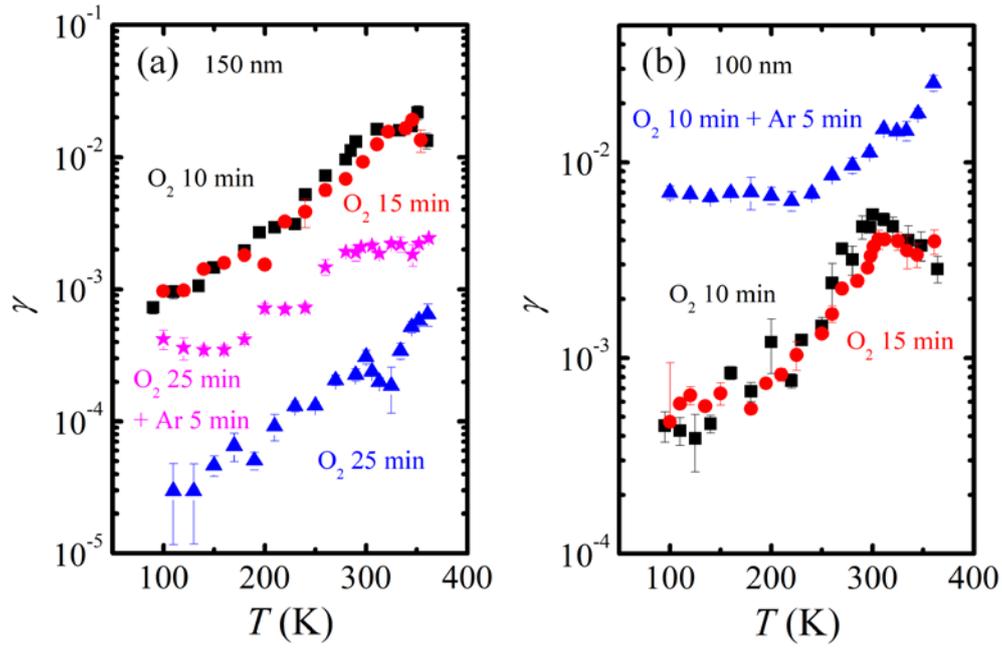

**FIG. 6.** (a) Variation of Hooge parameter $\gamma$ with temperature for four 150-nm-thick $RuO_2$ films underwent differing annealing conditions, as indicated. Annealing for 25 min in $O_2$ greatly reduces the $\gamma$ value, whereas an additional 5 min annealing in Ar significantly increases the $\gamma$ value. (b) Variation of $\gamma$ with temperature for three 100-nm-thick $RuO_2$ films underwent differing annealing conditions, as indicated. An additional annealing for 5 min in Ar significantly increases the $\gamma$ value.

**Figure 7**

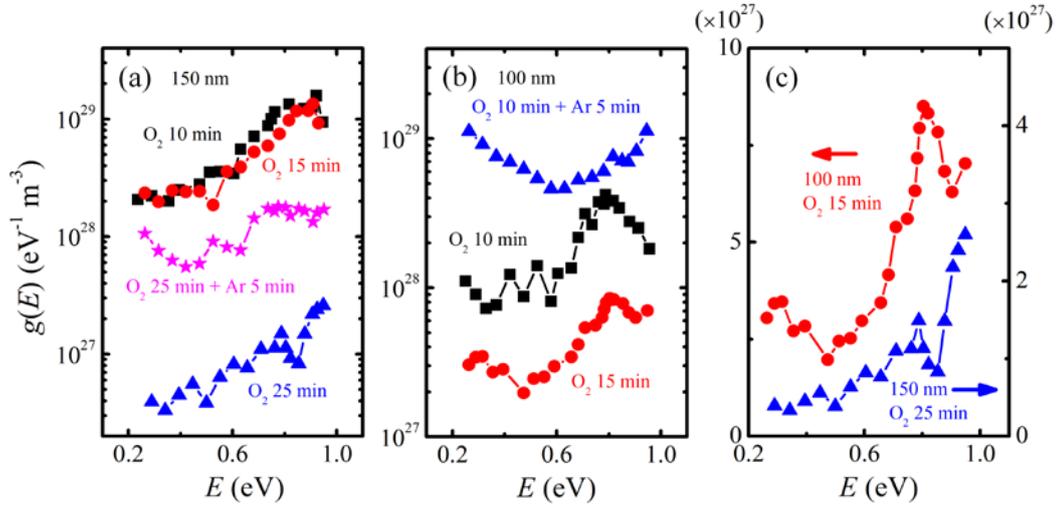

**FIG. 7.** Variation of calculated distribution function $g(E)$ with activation energy $E$ for (a) four 150-nm-thick films whose $\gamma$ values are plotted in Fig. 6(a), (b) three 100-nm-thick films whose $\gamma$ values are plotted in Fig. 6(b). (c) $g(E)$ for two representative $RuO_2$ films, as indicated. Note that $g(E)$ is plotted in linear scales to demonstrate a peak centering around $E \approx 0.8$ eV.